# Signatures of Kitaev interactions in the van der Waals ferromagnet VI₃

Yiqing Gu[1,2], Yimeng Gu[1], Feiyang Liu[1], Seiko Ohira-Kawamura[3], Naoki Murai[3] & Jun Zhao[1,2,4,5,*]

[1]*State Key Laboratory of Surface Physics and Department of Physics, Fudan University, Shanghai 200433, China*

[2]*Shanghai Research Center for Quantum Sciences, Shanghai 201315, China*

[3]*Materials and Life Science Division, J-PARC Center, Tokai, Ibaraki 319-1195, Japan*

[4]*Institute of Nanoelectronics and Quantum Computing, Fudan University, Shanghai 200433, China*

[5]*Shanghai Branch, Hefei National Laboratory, Shanghai 201315, China*

## Abstract

Materials manifesting the Kitaev model, characterized by bond-dependent interactions on a honeycomb lattice, can host exotic phenomena like quantum spin liquid states and topological magnetic excitations. However, finding such materials remains a formidable challenge. Here, we report high-resolution inelastic neutron scattering measurements performed on VI₃, a van der Waals ferromagnetic Mott insulator, covering a wide range of reciprocal space. Our measurements unveil highly anisotropic magnetic excitations in momentum space. Through a comprehensive comparative analysis of various models that incorporate diverse symmetry-allowed magnetic interactions, we find the observed excitations are well captured by a model with a large bond-dependent Kitaev interaction. These results not only help to understand the intriguing properties of VI₃, such as the pronounced anomalous thermal Hall effects and strong pressure or structure dependence of magnetism, but also open a new avenue for exploring Kitaev physics.

The Kitaev model has intrigued the scientific community with its description of bond-dependent interactions between spins on a honeycomb lattice, resulting in quantum spin liquid ground states or topological magnetic excitations [1-6]. This model has sparked widespread interest in the search for materials that exhibit Kitaev interactions [4,7-15]. One particular avenue of investigation involves the role of spin-orbit coupling in honeycomb magnets with bonds formed by edge-shared ligand octahedra, as it offers a promising pathway to realize the exotic phenomena predicted by the Kitaev model rooted in orbital spatial orientations [16]. Initially, research efforts were predominantly focused on $4d$ and $5d$ transition metal honeycomb magnets [8-14,17,18], with a special emphasis on magnetic ions such as $Ir^{4+}$ and $Ru^{3+}$ that possess strong spin-orbit coupling. Recent studies have expanded the exploration into $3d$ transition metal compounds, particularly those containing $Co^{2+}$ ions [15,19,20]. Nevertheless, the question of whether Kitaev interactions exist in cobalt magnets continues to be a topic of ongoing debate [21,22]. Furthermore, while most experimental investigations on Kitaev spin liquid materials have predominantly focused on antiferromagnetic systems exhibiting a zigzag order with $S = 1/2$, theoretical studies have suggested the potential existence of Kitaev spin liquids in higher-spin systems near ferromagnetic instabilities [23-26]. However, experimental studies in this direction have been scarce. Exploring the interplay between ferromagnetism and Kitaev physics may hold the potential to unlock new possibilities for the discovery and engineering of novel magnetic materials.

Recently, two-dimensional (2D) van der Waals (vdW) ferromagnets have attracted tremendous attention due to their potential for advancing 2D spintronic applications [27,28]. Among these materials, the newly discovered van der Waals ferromagnet $VI_3$ stands out for its distinct behavior compared to other systems like $CrI_3$ which possess nearly quenched orbital moment [29-34]. In $VI_3$, the electron configuration of $V^{3+}$ ions results in partially filled $t_{2g}$ orbitals, which makes spin-orbit coupling important for its magnetism. Indeed, neutron diffraction [32] and x-ray magnetic circular dichroism [33] measurements have revealed a substantial orbital moment in $VI_3$, with two V sites exhibiting different orbital occupations. Density functional theory (DFT) calculations support this observation by suggesting that the state with orbital-moment is energy favorable over the orbital-quenched state [33,35].

More surprisingly, the Curie temperature of $VI_3$ increases from 50 K to 60 K as the number of layers decreases towards the monolayer limit [36]. This contradicts the common expectation that lower dimensionality enhances fluctuations and lowers the magnetic transition temperature [27,28]. Furthermore,

VI$_3$ displays a large anomalous thermal Hall effect, indicating the potential existence of topological magnetic excitations [37]. These intriguing characteristics of magnetism in VI$_3$, combined with its remarkable sensitivity to structural distortion and external pressure [38,39], imply the presence of unusual magnetic interactions that are yet to be fully understood.

While previous studies utilizing inelastic neutron scattering in VI$_3$ have predominantly focused on investigating the magnetic excitations near the Brillouin zone center $\Gamma_1$, revealing two branches of dispersive excitations originating from two different V$^{3+}$ orbital states [40], the magnetic excitation spectra in the vicinity of the zone boundary and in the neighboring inequivalent zone near $\Gamma_1^*$, which hold crucial information for unraveling the underlying magnetic interactions, remain unclear. A comprehensive analysis of these magnetic excitation spectra in the unexplored regions is essential for a complete understanding of the magnetic interactions in VI$_3$.

Here, we report the inelastic neutron scattering measurements on VI$_3$ in a wide range of Brillouin zones (See Supplemental Material [41] and Refs. [42-44] for the details of experimental methods). The magnetic V$^{3+}$ ions form the honeycomb lattice with edge-sharing I$^-$ octahedral at room temperature in VI$_3$ [Fig. 1(a)]. As the temperature decreases, VI$_3$ undergoes a structural distortion at $T_{s1}$ = 79 K, followed by a ferromagnetic transition at $T_C$ = 50 K, as depicted in Fig. S6 in Supplemental Material [32,45-47]. With further cooling, a spin reorientation takes place ($T_{FM2}$ = 27 K), accompanied by an additional tiny lattice distortion. Considering the minimal lattice distortions observed, we approximate the lattice as a honeycomb structure at low temperatures. Figure 1(c) illustrates the high-symmetry points and directions in the reciprocal space. Black dashed lines indicate the Brillouin zone boundaries. There are two inequivalent zones, which are centered at $\Gamma_1$ $(1, 1, 0)$ and $\Gamma_1^*$ $(1, 0, 0)$ points, respectively.

Figure 2(a)–(f) presents the spin excitation spectra at various energies, covering a wide range of the Brillouin zone at 5 K. Given the rather weak dispersion of magnetic excitations along the $L$ direction (Fig. S1 in Supplemental Material [41]), the excitation spectra are integrated over -3 ≤ $L$ ≤ 3. It is observed that the spin excitations originating from the $\Gamma_1$ point disperse outward and form a six-pointed star shape pattern above 5.5 meV. Interestingly, the excitations arising from the zone center $\Gamma_1^*$ point exhibit a distinct pattern from that near the $\Gamma_1$ point and form a triangular-shaped pattern dispersing to the zone boundary.

To determine the detailed dispersion of the excitations, we present the excitation spectra in the $E$-$\mathbf{Q}$ space along the M-K-$\Gamma_1$-K-M and M-$\Gamma_1^*$-M directions. In addition to the low-energy mode below 6.5 meV observed in Figure 2, we have identified another high-energy mode above 6 meV [Fig. 3(a)&(c)]. The energy cuts reveal that the high-energy mode exhibits a significantly weaker intensity compared to the low-energy mode [Fig. 4(a)–(c)]. The low-energy mode displays a spin gap of approximately 3.9 meV, which reduces to about 2 meV when the temperature rises above $T_{\mathrm{FM2}}$. Meanwhile, the dispersion of the high-energy mode remains mostly unaffected (Fig. S2 in Supplemental Material [41]). This suggests that the low-energy mode corresponds to the spin wave of the ground state magnetic order, whereas high-energy mode is associated with the excited orbital state. This interpretation aligns well with the DFT calculations, which also indicate that the energy of the large-orbital-moment state is lower than that of the orbital-quenched state [33,35]. Consequently, the low-energy branch can be attributed to the ground state $a_{1g}e_-'^1$ [Fig. 1(d)], while the high energy mode is associated with the orbital-quenched $e_g'^2$ state [33,35] [Fig. 1(e)].

Furthermore, it is observed that the low-energy mode exhibits anisotropic V-shaped dispersion near $\Gamma_1^*$ [Fig. 3(c)]. Specifically, the dispersion is steeper on the high $H$ side in the $(H, 0, 0)$ direction, and a bend is observed at the low $H$ side at around 5.5 meV. These characteristics of the anisotropic dispersion are also consistently observed and quantitatively confirmed in the constant energy cuts which illustrate the transformation of a single peak into two peaks with increasing energy, followed by the further broadening of the peak on the left-hand side [Fig. 4(d)–(g)].

To understand the magnetic interactions responsible for the anisotropic spin excitation dispersion observed in VI$_3$, we performed a systematic exploration of various models with different magnetic interactions, such as the Heisenberg interaction, XXZ-type interaction, and symmetry-allowed Dzyaloshinskii-Moriya (DM) interaction. Our investigation revealed that they could not adequately capture the excitations observed in VI$_3$.

Given that VI$_3$ exhibits a considerable orbital moment [32,33], similar to 4$d$ or 5$d$ Kitaev magnets, such as $\alpha$-RuCl$_3$ [48], we recognized the need to incorporate anisotropic magnetic interactions such as the Kitaev and off-diagonal exchange interactions in the model. In materials featuring a honeycomb lattice and edge-shared ligand octahedra, the Kitaev interaction manifests as three kinds of bonds, each associated with bond-dependent Ising axes that are orthogonal to one another [Fig. 1(b)].

Remarkably, the incorporation of anisotropic magnetic interactions led to an effective description of the unusual excitation spectra observed in the material. The Hamiltonian of the $J$-$K$-$\Gamma$-$\Gamma'$-$A$ model, which we fitted to the excitation spectra of the low-energy magnon branch, is provided as follows:

$$H = \sum_{\langle i,j \rangle \in \alpha\beta(\gamma)} \left[ J \boldsymbol{S}_i \cdot \boldsymbol{S}_j + K S_i^\gamma S_j^\gamma + \Gamma\left(S_i^\alpha S_j^\beta + S_i^\beta S_j^\alpha\right) + \Gamma'\left(S_i^\alpha S_j^\gamma + S_i^\gamma S_j^\alpha + S_i^\beta S_j^\gamma + S_i^\gamma S_j^\beta\right) \right] + \sum_j A\left(\boldsymbol{S}_j \cdot \boldsymbol{n}\right)^2 \quad (1)$$

Here, $\langle i,j \rangle \in \alpha\beta(\gamma)$ denotes the type of nearest-neighbor bonds and the Ising axes of Kitaev interaction ($\gamma = x, y, z$ in local coordinates; $\alpha$ and $\beta$ are the other two orthogonal axes different from $\gamma$). $J$, $K$, ($\Gamma, \Gamma'$) and $A$ denote the Heisenberg interaction, Kitaev interaction, two types of off-diagonal exchange interactions and single-ion anisotropy, respectively. $\boldsymbol{n}$ denotes the direction of the easy axis, which is parallel to the magnetic moment direction of VI$_3$.

The simulated excitation spectra along the M-K-$\Gamma_1$-K-M and M-$\Gamma_1^*$-M directions are depicted in Fig. 3(b) and Fig. 3(d), respectively. Considering the unique in-plane magnetic structure of VI$_3$ (Fig. S6 in Supplemental Material), we took into account the presence of three magnetic domains, each characterized by different in-plane components of magnetic moments, separated by 120 degrees from one another, in our simulation. The V$^{3+}$ magnetic form factor was also considered.

To conduct these simulations, we employed the SpinW program based on the linear spin wave theory. The parameters that yielded the best fit for our simulations are as follows: $J$ = -1.04 meV, $K$ = -7.8 meV, $\Gamma$ = 0.3 meV, $\Gamma'$ = -1.2 meV, and $A$ = -0.6 meV. The results demonstrate good agreement between the simulated magnetic excitation spectra, obtained using this $J$-$K$-$\Gamma$-$\Gamma'$-$A$ model, and the experimental data along high symmetry directions in reciprocal space [Fig. 3(a)–(f)]. We note that the bend or inflection observed in the dispersion of the low-energy mode along the M-$\Gamma_1^*$-M direction is also captured by our model [Fig. 3(c)&(d)]. When considering the magnetic excitation dispersion along the $\Gamma_3$-M-$\Gamma_1^*$-M-$\Gamma_2^*$-M-$\Gamma_2$ path, the simulated excitation spectra [Fig. 3(f)] further reinforce the agreement with the experimental data [Fig. 3(e)]. In comparison, alternative models fail to explain the observed asymmetric dispersion for the low energy mode (Fig. S3&S4 in Supplemental Material [41]). In addition to the anisotropic low energy mode, the high-energy mode above 6.5 meV originating from the orbital-quenched state was simulated by the $J$-$A$ model considering the nearest-neighbor Heisenberg interaction and the single-ion anisotropy [Fig. 3(b), 3(d), 3(f)].

Furthermore, the triangular-shaped momentum structure of the low-energy magnetic excitations near the zone center $\Gamma^*(1,0,0)$ [Fig. 2(a)–(f)] exhibits remarkable concordance with the simulation of the $J$-$K$-$\Gamma$-$\Gamma'$-$A$ model [Fig. 2(g)–(l)]. In contrast, alternative models show a distinct ring-like pattern (Fig. S5 in Supplemental Material [41]). Our simulations incorporating the Kitaev interaction also replicate the distinctive six-pointed star shape pattern observed near another zone center at $\Gamma_1$ (1,1,0).

Nonetheless, we have observed some intensities filling in the center of $\Gamma_1$ (1,1,0) [Fig. 2(f)]. These intensities may arise from magnon damping or the presence of continuum excitations at the center, phenomena not explicitly accounted for in simple linear spin wave theory calculations. Interestingly, a similar six-pointed star-shaped scattering pattern with intensities filled in the center was also reported in Kitaev spin liquid material $\alpha$-RuCl$_3$ near the zone center $\Gamma_1$ [13,14]. In the case of $\alpha$-RuCl$_3$, the center continuum excitations are attributed to fractionalized Majorana fermion excitations, which could be linked to the observed thermal Hall effect [17].

The combination of our inelastic neutron scattering data and simulations presents compelling evidence for the existence of Kitaev interaction in VI$_3$, shedding light on the understanding of its other unusual behaviors. Particularly, VI$_3$ exhibits a significant anomalous thermal Hall effect, which is an order of magnitude higher than other ordered magnets [37]. This effect was previously attributed to topological spin excitations induced by DM interactions [37]. However, our findings suggest that the dominant Kitaev interaction dictates the low-energy Hamiltonian of VI$_3$, which could also induce anomalous thermal Hall effect [5,6,49]. This distinguishes VI$_3$ from orbital-quenched CrI$_3$, where DM interactions play a pivotal role [34]. The presence of Kitaev interactions in VI$_3$ may also account for the anisotropic thermal dynamic properties within the *ab*-plane [46], as opposed to the isotropic thermal dynamic properties usually resulting from conventional magnetic interactions [22].

Another intriguing phenomenon is the anomalous increase in the transition temperature ($T_C$) of VI$_3$ in the monolayer limit [36]. This contradicts DFT calculations based on Heisenberg-type interactions and anisotropy, which predict that the $T_C$ of the monolayer should be half of its bulk value [50]. Interestingly, a similar behavior of enhanced magnetic ordering temperature was also observed in the monolayer Kitaev magnet $\alpha$-RuCl$_3$ [51]. These collective findings point to the notable role that Kitaev interactions may play in the monolayer form. To gain a deeper understanding into this phenomenon, further calculations that fully consider the predominance of Kitaev interactions are required. Recently, evidence of Kitaev interactions was also found in geometrically frustrated triangular-lattice quasi-2D vdW magnets FeI$_2$ and CoI$_2$ [52-54], it will be interesting to explore whether similar phenomena occur in the monolayer forms of these systems.

Unlike intensively studied Kitaev spin liquid materials, such as $\alpha$-RuCl$_3$ and Na$_2$IrO$_3$, with the effective $S = 1/2$ spins and zigzag antiferromagnetic order [8,12,55,56], VI$_3$ possesses the effective $S = 1$ spins, since the 3-fold degeneracy of two-electron orbital states is lifted due to the presence of the trigonal distortion and spin-orbit coupling [Fig. 1(d)] [33,35]. Although the $S = 1$ Kitaev model is not exactly solvable like its $S = 1/2$ counterpart, theoretical and numerical works have shown that it can lead to a Kitaev spin liquid phase even in the presence of some non-Kitaev interactions [23,25,26]. For instance, Fukui *et al.* [26] proposed a parameter range for the $S = 1$ Kitaev spin liquid phase to exist, roughly $0.7150 \lesssim \xi \lesssim 0.7775$. Interestingly, VI$_3$ falls within this range, with a parameter value of $\xi = 0.7289$, situated near the ferromagnetic phase boundary. Here we denote $K = \sin(2\pi\xi)$ and $J = \cos(2\pi\xi)$. Similar phase diagrams have also been obtained by other calculations [25]. These calculations were conducted within the framework of Kitaev-Heisenberg model, without taking into account off-diagonal exchange interactions and single-ion anisotropy. When these factors are considered, they tend to push the system toward the ferromagnetic phase [57,58]. This intriguing observation hints at the role of frustration induced by Kitaev interactions as the key factor behind VI$_3$'s magnetism tunability under external pressure and strain [38,39]. Consequently, it opens up exciting possibilities for inducing the Kitaev spin liquid state and investigating its associated novel phenomena using various tuning methods. These methods include the application of magnetic fields, strain engineering, pressure control, and chemical substitutions on heavy ligands of the anion, which can effectively modulate the strength of spin-orbit coupling and Kitaev interaction [24]. These results position vanadium magnets as a new platform for the exploration of Kitaev physics, thereby expanding the field for exploring emerging quantum phenomena.

We thank W. Li, C. S. Xu, G. Chen, H. Wu, Y. Qi, H. Li, K. Yang and W. Q. Zhu for helpful discussions. This work was supported by the was supported by the Innovation Program for Quantum Science and Technology (Grant No. 2024ZD0300103), Key Program of National Natural Science Foundation of China (Grant No. 12234006), National Key R&D Program of China (2022YFA1403202) and the Shanghai Municipal Science and Technology Major Project (Grant No. 2019SHZDZX01). The neutron experiments at AMATERAS (BL-14) were carried out under approval of the Materials and Life Science Experimental Facility of the J-PARC (Proposal No. 2020B0202 and 2022B0114).

[*]zhaoj@fudan.edu.cn

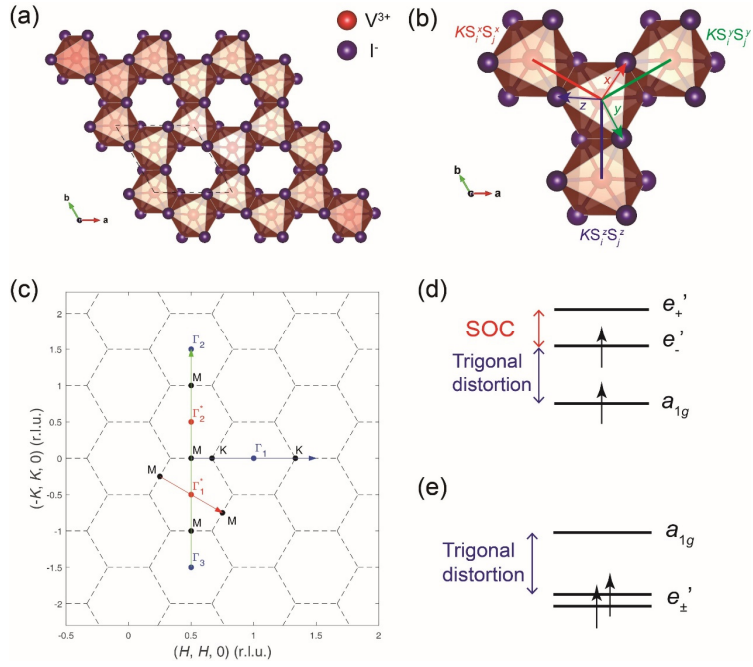

FIG. 1. Lattice structure, bond geometry of Kitaev interactions and schematics of crystal field splitting in VI$_3$. (a) Lattice structure of VI$_3$. The V$^{3+}$ and I$^-$ ions are colored in red and purple, respectively. (b) Bond-dependent Kitaev interactions in the local coordinates of VI$_3$. The arrows marked with $x$ (red), $y$ (green) and $z$ (blue) denote the Ising axes of corresponding V—V bonds, which are classified into $x$-type (red), $y$-type (green) and $z$-type (blue) bonds. (c) Reciprocal lattice and high symmetry directions. (d) The ground state orbital occupation of two $3d$ electrons ($a_{1g}e'^1_-$). The 3-fold degeneracy of the $t_{2g}$ orbital is lifted by the trigonal distortion and spin-orbit coupling. (e) The high energy orbital-quenched state with $e'^2_g$ occupation.

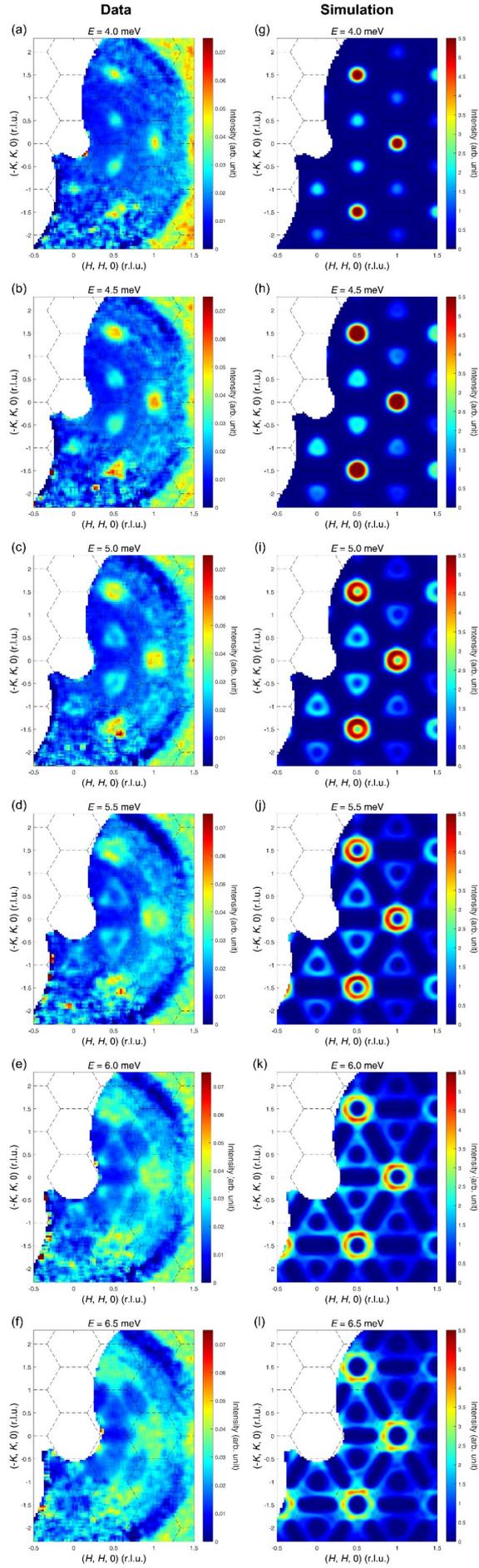

FIG. 2. Constant energy slices of magnetic excitations in VI$_3$ within the $(H, K)$ plane at $T = 5$ K. (a)–(f) Measured

constant energy slices with energy transfer $E$ = 4.0, 4.5, 5.0, 5.5, 6.0, 6.5 meV which are integrated over $E \pm 0.5$ meV and $-3 \leq L \leq 3$. The incident neutron energy is $E_i$ = 15.1 meV. (g)–(l) Constant energy slices at the specified energies generated through simulations using the $J$-$K$-$\Gamma$-$\Gamma'$-$A$ model.

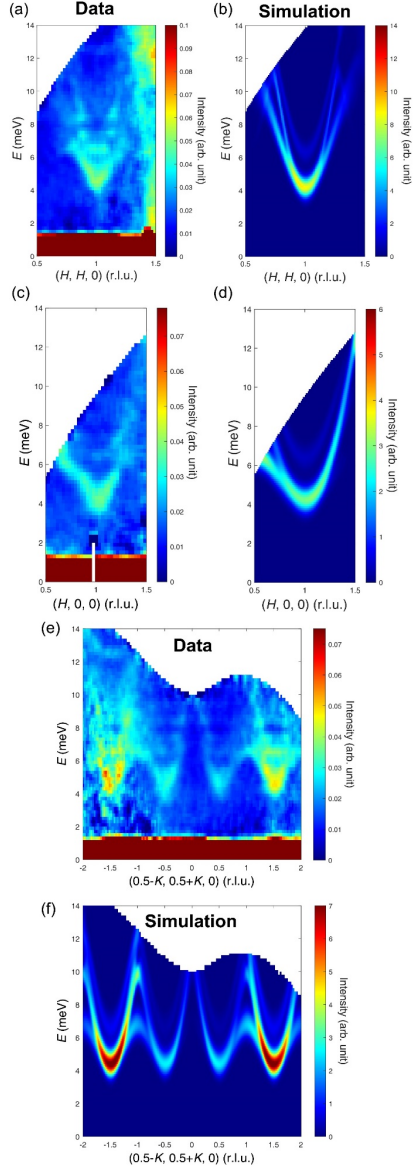

FIG. 3. Dispersion of magnetic excitation spectra of VI$_3$ at 5 K. (a) The momentum-dependent magnetic excitations along the M-K-$\Gamma_1$-K-M path. (b) Simulated magnetic excitations along the M-K-$\Gamma_1$-K-M path. The low-energy branch, originating from the high-orbital-moment state, is calculated using the $J$-$K$-$\Gamma$-$\Gamma'$-$A$ model with parameter specified in the main text. The high-energy branch, excited from the orbital-quenched state, is calculated using the parameters with $J$=-2.9 meV, $A$=-3.1 meV. Note that the magnetic exchange parameters associated with the high energy mode cannot be unambiguously determined based on the available data. (c) The momentum-dependent magnetic excitations along the M-$\Gamma_1^*$-M path at 5 K. A distinct bend anomaly is observed at $\sim$ 5.5 meV in the low-energy branch. (d) Simulated magnetic excitations along the M-$\Gamma_1^*$-M path. (e) Magnetic excitation spectra measured along the $\Gamma_3$-M-$\Gamma_1^*$-M-$\Gamma_2^*$-M-$\Gamma_2$ path at 5 K. (f) The simulated magnetic excitation spectra along the $\Gamma_3$-M-$\Gamma_1^*$-M-$\Gamma_2^*$-M-$\Gamma_2$ path.

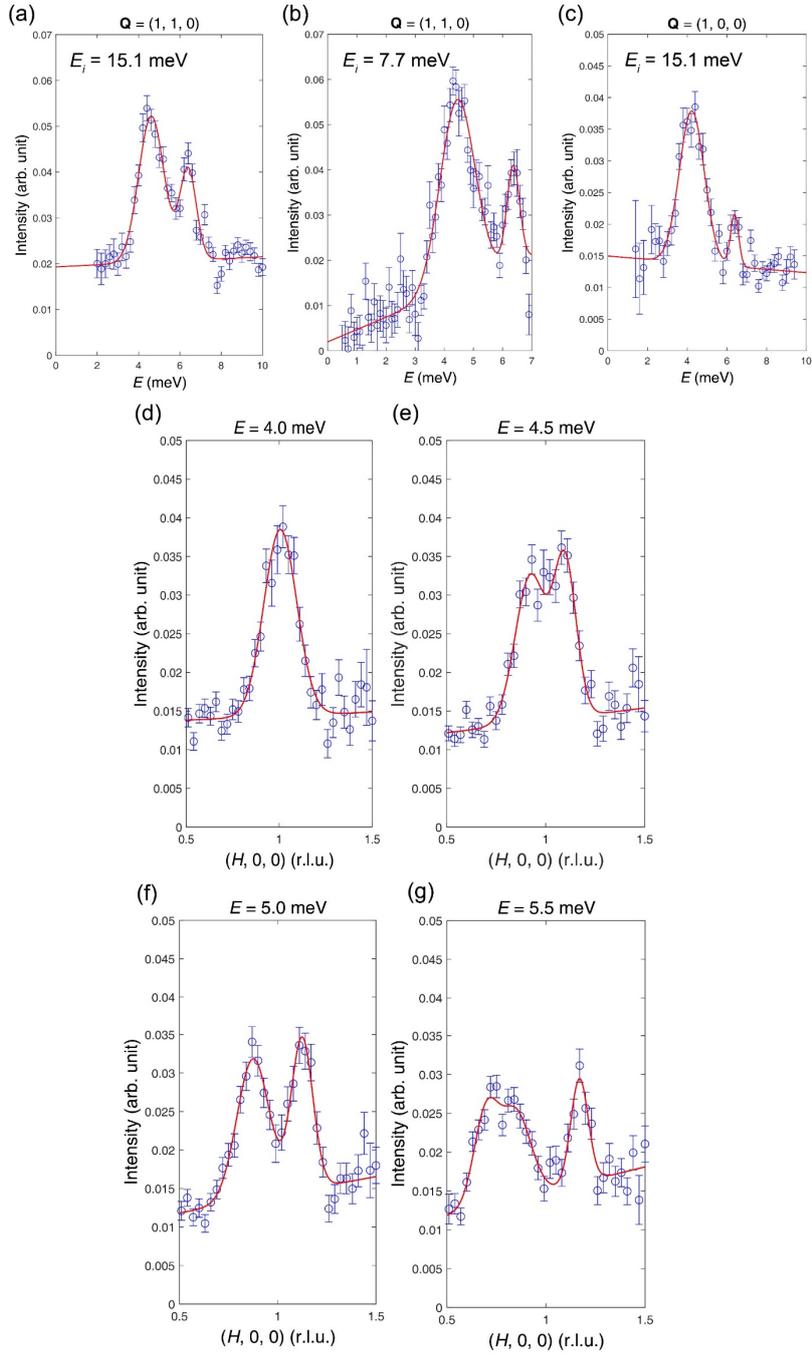

FIG. 4. Constant momentum (**Q**) and constant energy cuts of VI$_3$ magnetic excitation spectra. (a)–(b) Constant **Q** cuts at $\Gamma_1$ $(1, 1, 0)$ obtained from the 5 K data with incident neutron energy $E_i$ = 15.1 meV and 7.7 meV, respectively. (c) Constant **Q** cuts at $\Gamma_1^*$ $(1, 0, 0)$ obtained from the 5 K data with $E_i$ = 15.1 meV. (d)–(g) Constant energy cuts along the M-$\Gamma_1^*$-M path with energy transfer $E$ = 4.0, 4.5, 5.0 and 5.5 meV, respectively. The constant **Q** and constant energy cuts are fitted by Gaussian profiles with linear background. The error bars indicate 1 standard deviation.